# Modelling of Sickle Cell Anemia Patients Response to Hydroxyurea using Artificial Neural Networks

Brendan E. Odigwe[1], Jesuloluwa S. Eyitayo[2], Celestine I. Odigwe[3], Homayoun Valafar[1]

[1]Department of Computer Science & Engineering, University of South Carolina, Columbia, SC, USA.

[2]Department of Computer of Science, Texas State University, San Marcos, Texas, USA.

[3]Department of Internal Medicine, University of Calabar Medical School, Calabar, C.R.S, Nigeria.

**Abstract**

*Hydroxyurea (HU) has been shown to be effective in alleviating the symptoms of Sickle Cell Anemia disease. While Hydroxyurea reduces the complications associated with Sickle Cell Anemia in some patients, others do not benefit from this drug and experience deleterious effects since it is also a chemotherapeutic agent. Therefore, to whom, should the administration of HU be considered as a viable option, is the main question asked by the responsible physician. We address this question by developing modeling techniques that can predict a patient's response to HU and therefore spare the non-responsive patients from the unnecessary effects of HU. We analyzed the effects of HU on the values of 22 parameters that can be obtained from blood samples in 122 patients. Using this data, we developed Deep Artificial Neural Network models that can predict with 92.6% accuracy, the final HbF value of a subject after undergoing HU therapy. Our current studies are focusing on forecasting a patient's HbF response, 30 days ahead of time.*

Key words: Artificial, Neural, Network, Deep, Sickle, Cell, Anemia, predictive.

## 1. Introduction

Sickle cell anemia is an inherited form of anemia, a condition in which there is not enough healthy red blood cells to carry adequate oxygen throughout the body. Red blood cells contain the oxygen-carrying molecule called Hemoglobin (Hb) which are of different types depending on the composing sub-units. Sickle cell disease patients have the Sickle Hemoglobin (HbSS) sub-unit, which does not carry the amount of oxygen needed. Fetal Hemoglobin (HbF) is another sub-unit that is normally lost as we age but is not subject to the mutation that causes Sickle Cell Anemia (SCA). Hydroxyurea has been observed to increase the amount of HbF in circulation. Most patients respond to Hydroxyurea (HU) therapy with an increase in the HbF concentration of blood [1], therefore reducing the number of incidents related to Sickle Cell Anemia. Although Hydroxyurea, in general, reduces the number of complications of the SCA disease in most patients, it has numerous deleterious effects since it is an agent used during chemotherapy. Therefore, the primary challenge facing a physician who is providing care to an SCA patient is whether the benefits of the HU outweigh its detriments. In the absence of any prior knowledge, a trial process of more than six months is initiated in order to determine a patient's response to potentially a harmful treatment.

Pattern recognition and value prediction techniques based on the employment of artificial neural networks have had great success especially in establishing complex relationships and identifying patterns between disparate parameters such as age, gender or genetic information. Artificial Intelligence tools, including Deep Neural Networks (DNN), have historically been used in the development of Clinical Decision Support [2], medical image processing [3][4], and prediction of patient outcomes [5][6].

Availability of predictive tools can be of paramount importance in the development of personalized medicine. More specific to the SCA, if the magnitude of the HU-elicited increase in the percentage fetal hemoglobin in a patient with sickle cell anemia can be predicted, it will aid in the identification of responders and non-responders to the therapy, non-compliant patients, and an estimation of the extent of response for the identified responders [7]. This approach could also give physicians the incredibly important ability of predicting whether a patient's sickle cell symptom(s) will be significantly reduced by the therapy. Having the ability to look at a patient on an individual basis and being capable of determining if the projected benefits of treatment are worth the deleterious side effects that could ensue, will allow for a more specific patient guided therapy.

Previous work [5] reported the utility of ANN in predicting the effectiveness of Hydroxyurea therapy for Sickle cell anemia patients. In that report, a shallow ANN was used to identify HU treatment responders using a limited definition of responders. Here we continue our previous work by establishing the use of ANN in a more pragmatic definition of responders as patients who more than double their initial %HbF. We also perform additional data analytics in order to expose the challenging nature of this dataset. Finally, we provide a preliminary report on the utilization of LSTM [8] to predict a patient's response to HU one month in advance.

## 2. Background

Sickle Cell Disease/Anemia is an autosomal recessive genetic disease where Red Blood Cells (RBC's) take the shape of a crescent/sickle. Sickle cell mutation is a non-conservative missense mutation where the Glutamate (hydrophilic) in the 6th amino acid of the beta globin is substituted with a Valine (hydrophobic). A hemoglobin tetramer which is 2- alpha and 2-mutated beta is called Sickle Hemoglobin (HbS). HbS is able to carry oxygen, but when deoxygenated, it changes its shape, making it capable of clustering with other HbS complexes that form polymers and distort the shape of red blood cells into a crescent shape (Sickling)[9].This change allows the easier destruction of erythrocytes, causing anemia, among other things [10].

Hydroxyurea (Hydrea or HU) is a preventive medication that reduces the complications associated with Sickle Cell Anemia (SCA). HU operates by increasing the amount of gamma-globin in the RBCs which results in the expression of more fetal hemoglobin (HbF) that is not subject to Sickle Cell mutation. HbF is the primary Hemoglobin during the fetal development stages and at birth, which explains why Sickle cell symptoms do not appear until a few months of life (when adult Hb, containing the mutated beta globin starts to dominate) [11].

As with all medications, positive response is not guaranteed; some patients respond positively depicted by an increase in the concentration of their fetal hemoglobin, while other patients do not respond positively (see fig 1). While Hydrea reduces the complications associated with SCA, it is a chemotherapeutic agent and may elicit the following side effects: nausea, vomiting, diarrhea or constipation, acute pulmonary reactions, genetic mutation, secondary leukemia, and hair loss, to name a few [12] . It is therefore beneficial to spare the non-responding patients from the unnecessary exposure to this drug's side effects.

In previous work [5], ANNs have been utilized to predict SCA patients responsiveness by way of a classification model, which categorizes patients into responders and non-responders. the criteria employed for classification of responders consisted of patients whose HbF concentration exceeded the 15% threshold. Based on this definition of a responder, a person whose %HbF increased from an initial value of 14.9 to a final value of 15.1 is considered a responder. In contrast, a person whose %HbF increased from an initial value of 1.0 to a final value of 14.8 is considered a non-responder. Based on similar observations, the community has scrutinized the definition of a responder, and new definitions have been proposed. In this work, we explore the usability of ANN as a predictive tool for the early identification of responders based on different criteria. In addition, we present a new approach in utilizing the predictive power of Machine Learning techniques that allows us to remain agnostic to the definition of a responder.

# 3. Materials and/or Methods
## 3.1. Sickle Cell Anemia Patient Data
In this experiment, data from 122 sickle cell anemia patients were collected over a varying period, depending on the patient's response to the treatment and adherence to the project protocol. The cohort of patients were at least 16 years of age, who were treated with a daily, oral dose (based on body weight) of Hydroxyurea with subsequent increase in dosage if the HbF levels stabilized in successive monthly visits. Only patients who had received HU for 8 months or longer were included in the ANN experiments, although, certain patients began to display positive response well after 8 months of continuous treatment.

Blood samples were obtained and analyzed monthly. The values of 22 parameters [5] for each of the 122 patients were recorded for a minimum of 8 months and a maximum of 98 months. The parameters analyzed for each patient are those normally included in a Complete Blood Count (CBC), Chemistry Profile (SMA18) serial HbF levels (absolute and percent), DNA analyses for globin gene haplotype and the number of genes, treatment duration, weight, age, and gender. The values of these parameters were used to train ANNs and for statistical analyses [5].

## 3.2. Statistical Analyses
Our initial step in understanding the complexity of this study consisted of performing several standard analytics of the data. In addition to performing the first and second-ranked statistical analyses, density estimation, principal component analysis, and correlation coefficient analysis were performed on the entire dataset.

Density estimation [13] aims to visualize the effects of HU on patients who undergo the therapy by observing the distribution of HbF before and after the therapy across the entire cohort of participants.

Principal component analysis [14] aims to explore meaningful strategies in performing dimensional-reduction in the principal component space to remove the unnecessary complexities of the problem. Correlation coefficient [14] , on the other hand, will be useful in reducing the dimensionality of the problem in the original parameter space by identifying the data with degree of correlation.

## 3.3. Artificial neural networks approach
This project utilized Artificial Neural Networks [15] (ANN) as the primary predictive Machine Learning approach. Our utilization of ANNs as the core predictive tool is based on the minimalist approach of employing the simplest model that will satisfy the problem requirements. Therefore, our investigations utilized a spectrum of ANN architectures including the legacy shallow ANNs. Our developments included neural network models using MATLAB [16]; or TensorFlow [17] and Keras [18].

Within the dataset, some patients did not have complete values for all 22 parameters. In this work, the missing data were substituted with a zero since it has no contribution to the updated weights during the learning process of the back-propagation algorithm. A Shift of one was added to all data with a legitimate value of 0 (e.g. number of BAN). In addition, the dataset contained data with substantially different ranges. For instance, the age of the patient is measured in days (range was in thousands), while other parameters such as haplotypes units have single digit values. This disparity in the range of data is automatically resolved by normalization of data in MATLAB during the training/testing processes.

Due to the limited amount of data, we employed an n-fold cross validation approach; train with *n-1*, test with *1*, repeat *n* times, where *n* is the total number of patients in the dataset. This approach was applied for all experiments conducted in this study. The details related to the optimal network architecture and the training-testing processes are discussed in the individual results sections.

## 3.4. Specific aims of the project

After data preprocessing, we explored the development of neural network models to carry out the following three aims.

### 3.4.1. Aim 1: Classify responders and non-responders.
We developed a shallow two-stage fully connected, feed-forward artificial neural network topology in MATLAB with a backpropagation learning algorithm for training and testing.
Patients whose post-therapy HbF experienced a 15% increase or more were classified as responders and anyone whose values did not meet that threshold were classified as non-responders. This was performed as a repetition of earlier work[19], [20] to establish comparability in results and further ensure the integrity of results. 72 of the total 122 patients with the highest number of observations were used for this experiment.

Due to the lack of general agreement on the specific percentage regarded as responding, we established a threshold for a positive response as patients who experience a HbF increase of 100% (double the initial HbF) or more because of its popularity among the community of Sickle Cell Anemia (SCA) researchers.

### 3.4.2. Aim 2: Predict final fetal hemoglobin level.
After the initial results obtained from implementing a shallow neural network in MATLAB for classifying responders and non-responders, the next step is to attempt predicting a patient's final fetal hemoglobin value after undergoing Hydroxyurea therapy. This is to predict the highest %HbF observed over the course of medication administration.

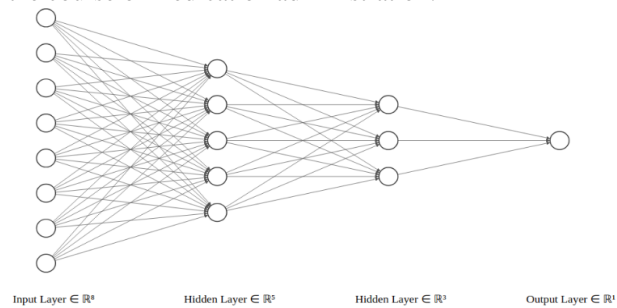

*Figure 1. Diagram of a feedforward Artificial Neural Network*

### 3.4.3. Aim 3: Prediction of %HbF one month ahead.
Although it may be of theoretical benefit to gain prior knowledge of a patient's maximum response to HU therapy (from Aim 2), such information is of little pragmatic use without knowing when the effect will be observed (in 6 months or 6 years of therapy). To resolve this limitation we have explored a new approach where we can predict a patient's %HbF one month in advance as a preliminary step toward predicting the time when the full effect of the medication will be observed. To that end, we implemented a recurrent neural network (RNN) with Long short-term memory (LSTM) for the time series prediction.

A recurrent neural network is a class of artificial neural network where connections between nodes form a directed graph along a temporal sequence. LSTM networks are well-suited to making predictions based on time series data, since there can be lags of unknown duration between important events in a time series, as is the case with the problem at hand [21], [22].

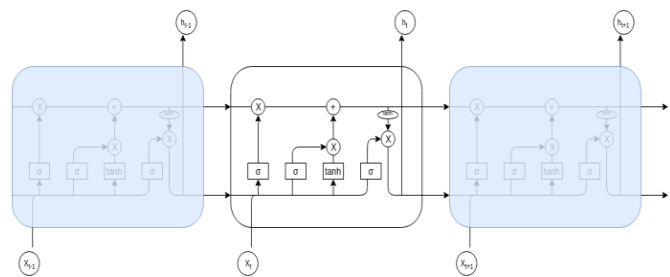

*Figure 2: Diagram of an LSTM Neural Network*

Regular feed-forward and recurrent deep neural network were developed using TensorFlow (an interface for expressing and executing machine learning algorithms) and Keras (Python machine learning application programming interface, which allows you to easily run neural networks). Training and testing were performed using the patient data previously obtained after pre-processing. The neural network architecture was regularly modified to obtain the optimal model.

The neural network model used for the final HbF prediction was a deep model constituting 3 hidden layers. 21 input neurons, 12 neurons for the first hidden layer, 8 neurons for the second hidden layer, 4 neurons for the third hidden layer and then a single neuron at the output layer (21-12-8-4-1) since just the single value prediction is expected.

We made use of the Adam optimizer for iterative update of the network weights and for each layer within the model, we made use of the rectified linear unit (ReLU) activation function with exception to the final layer which has a linear activation function because the output is expected to be a continuous value. This network architecture was used to predict the final fetal hemoglobin value after treatment.

The neural network model used for the periodic prediction was a recurrent LSTM model. The data was prepared by normalizing the features and framing the dataset as a supervised learning problem as predicting the HbF response at the current month given the response and patient biological data from the previous month. The first input layer constitutes 50 neurons and 1 neuron in the output layer. The input shape is 1 time-step with 22 features, because the input layer receives the 21 parameter values of the current time-step and the resultant prediction of the previous time-step.

## 4. Results and discussion

### 4.1. Statistical analyses

As the first step in conveying the complexity of the problem, Figure 3 illustrates some examples of patient responses to the HU treatment. As illustrated in these figures, not all patients' response to HU results in a monotonically increasing levels of HbF. Furthermore, it is evident that in some instances, an initial period of response follows by a decline in the HbF levels. Visual inspection of these response profiles, demonstrates the complexity of predicting a patient's response to HU treatment.

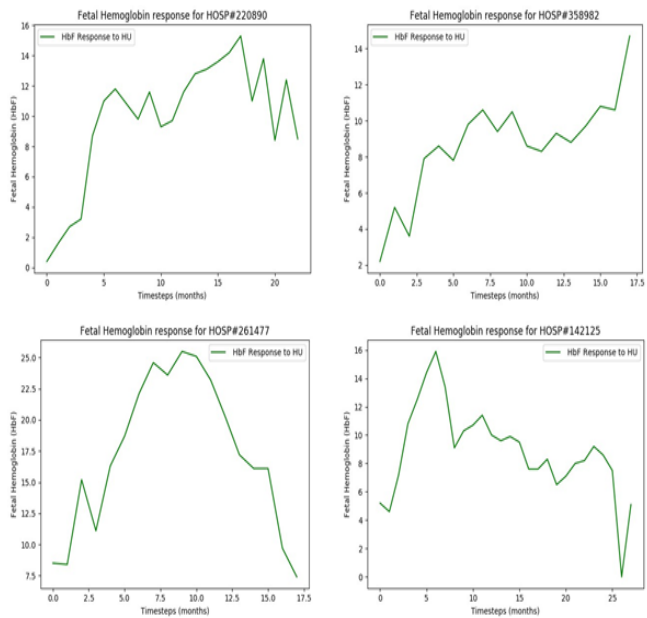

*Figure 3. Plots showing the varying response of different patient's Fetal Hemoglobin to Hydroxyurea*

Application of density estimation provides supporting evidence for the effectiveness of HU. Figure 4 shows the distribution of %HbF across all patients before (depicted with a continuous line) and after (depicted with a dashed line) the Hydroxyurea therapy. The vertical lines (in Figure 4) correspond to the mean values for each distribution. It is evident from this figure that there is a clear increase in the %HbF for most patients. In fact, the figure shows a notable increase in the average HbF value of the patients in the population. Furthermore, the distortion in the distribution of the %HbF before and after the HU therapy, indicates that patients exhibit an unequal response to the drug. One-tailed student T-test was applied to the results of statistical analyses to confirm that the observed variation in concentration of HbF, as well as other results, were due to the HU administered and not due to random events. A confidence level of 0.001 was the minimum value accepted as significant in the results of Student T-tests.

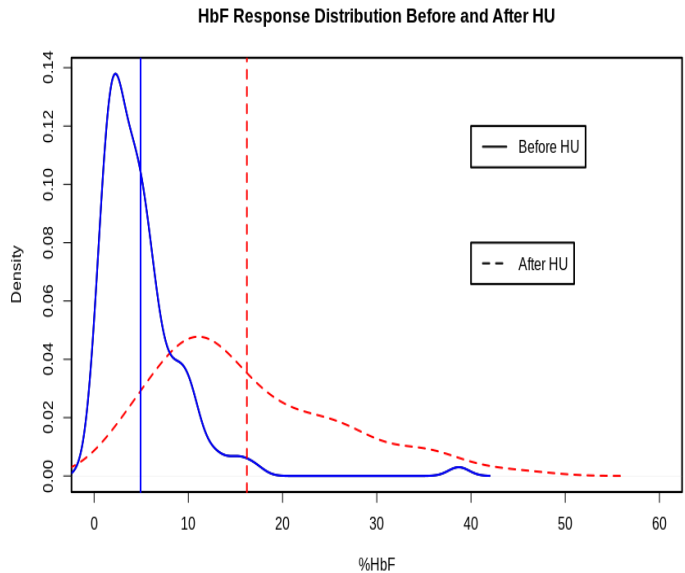

*Figure 4. Density distribution of SCA patients %HbF before and after undergoing hydroxyurea therapy*

The correlation coefficient analysis results are shown in the heat map (shown in Figure 5). As shown in this figure, the 22 parameters of this problem show correlation values within the mid range of -0.5 to 0.5, indicating that there is no substantial correlation between any two parameters in the dataset to assist with dimensional reduction.

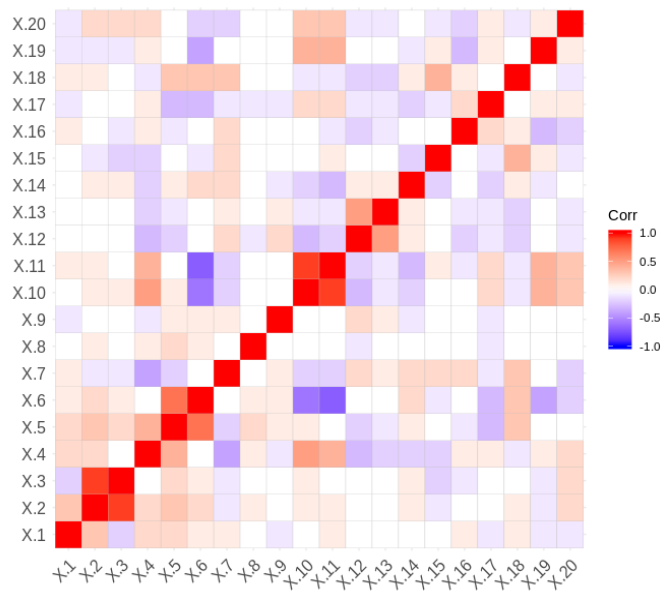

*Figure 5. Correlation coefficient heatmap of all parameters in the patient dataset*

Finally, the results of the spectral analysis obtained from principal component analysis are summarized in Figure 6. Based on the gradual decrease in the spectral analysis, it can be concluded that dimensional reduction in the principal component space is not a very effective approach.

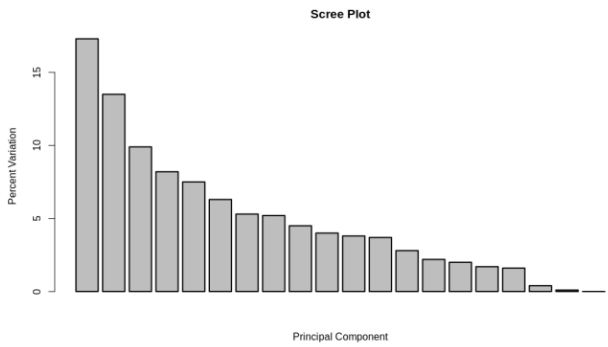

*Figure 6. Spectral analysis from Principal Component Analysis conducted on the patient dataset*

In summary, the collective results of the statistical analyses exposes the complexity of the problem statement and suggests that the problem could benefit from the application of artificial neural networks.

### 4.2. Results for Aim 1

The neural network that was utilized during this experiment consisted of a single hidden layer with 4 hidden neurons. This ANN was able to identify responders and non-responders based on the previously reported 15% threshold, with 83 percent accuracy. Although redundant, the repeat of the previous results was necessary to establish equality and continuity between the current and past work. A total of 72 patients were used for this study and 60 of them were accurately classified by the neural network.

The classification experiment was repeated with the same ANN topology using the doubling of %HbF criterion as the identification of responders. During this experiment, 98 of 122 patients were predicted correctly producing a final accuracy of 80.5%. Specificity (the percentage of responders which were actually predicted as responders) was 88.5% and sensitivity (the number of responders which were predicted to be non-responders) was 60%. In general the performance of this network indicated a slight bias towards prediction of responders.

### 4.3. Results for Aim 2

Predicting the final value of a patient's HbF concentration allows the data analysts to avoid the lack of consensus within the community of physicians and caregivers. In this mode, we can simply predict the final %HbF value that a patient will achieve, and allow the physician's to justify the use of HU. Restatement of the problem turns the problem definition from a classification problem to regression problem. The model will be predicting a real continuous value and this experiment was performed on the total dataset comprising of 122 patients. During this study we utilized a neural network with 3 hidden layers and 24 (12-8-4) total neurons , which was modelled with TensorFlow and Keras.

Experimentation was carried out using all 122 patient data for training and testing. During the evaluation step, we used a percentage error threshold of 30% between the predicted value and the actual value. That is, if the actual value is 10 and the predicted value (P) is greater than or equal to 7 and less than or equal to 13, it is considered correct, else, the prediction is considered false. With this approach to evaluation, 113 of the 122 predictions were correct giving a 92.6% accuracy level.

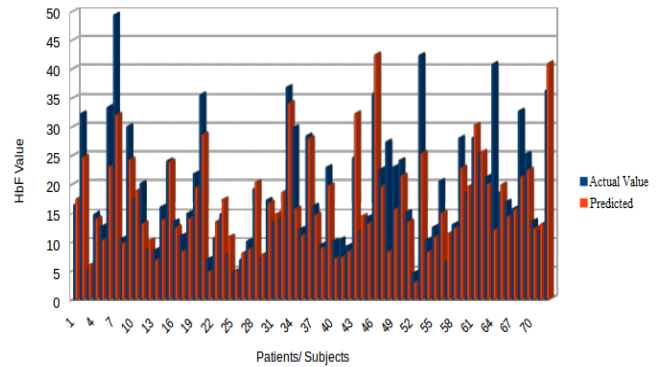

*Figure 7: Predicted HbF value and their corresponding Actual Values*

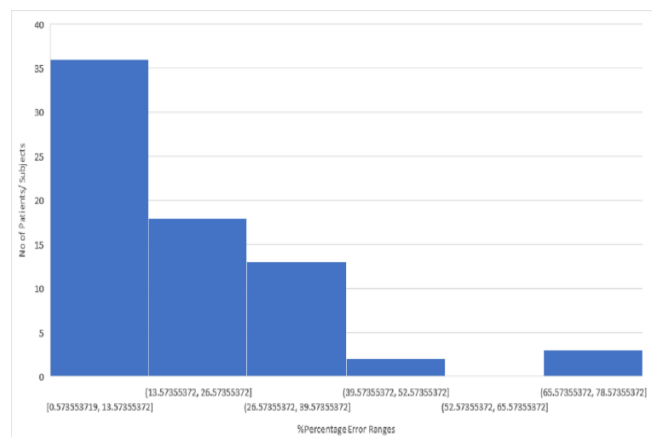

*Figure 8: Percentage Error Distribution of all ANN results*

### 4.4. Results of Aim 3

Although the reformulation of the problem from classification into regression domain alleviates the dependency on the accepted definition of a responder, it does not remove other existing logistical limitations. For instance, while the regression approach may predict correctly the maximum achievable value of %HbF, it fails to provide the conditions under which such outcomes can be achieved. Some of the critical values that may be required in this prediction are the drug dosage and the time needed for the HU therapy to take the maximum effect. Here we present the results of a preliminary approach that allows prediction of

%HbF for a given patient 30 days ahead of time. In this exercise we employed the use of LSTM networks (modelled in TensorFlow and Keras) on a small subset of patients. This was performed as a series of regression problems (depending on the number of months the patient underwent therapy) with continuous output values at each step.

For each patient, we used a series of expected values reflecting the next month's HbF value. The monthly predictions are expected to follow the same sequence of progression or decline to be considered a successful prediction.

The Pearson's correlation coefficient [14] (PCC) was used as a metric to determine the similarity in the monthly progression of the patient's predicted and the expected HbF values. The results corresponding to our six preliminar patients are shown in Figure 9. In this figure, trends illustrated in red and green correspond to the actual and the predicted response trajectories of the six test patients respectively. Four of these patients exhibited PCC between 0.6-1, 1 patient exhibited PCC in the intermediate range of 0.3-0.59, and the remaining patient exhibited poor PCC score (less than 0.29).

The training process consisted of utilizing monthly observations across 121 subjects. The testing process consisted of obtaining the next predicted value of HbF for the excluded patient over the course of their therapy. This process was repeated for 6 patients.

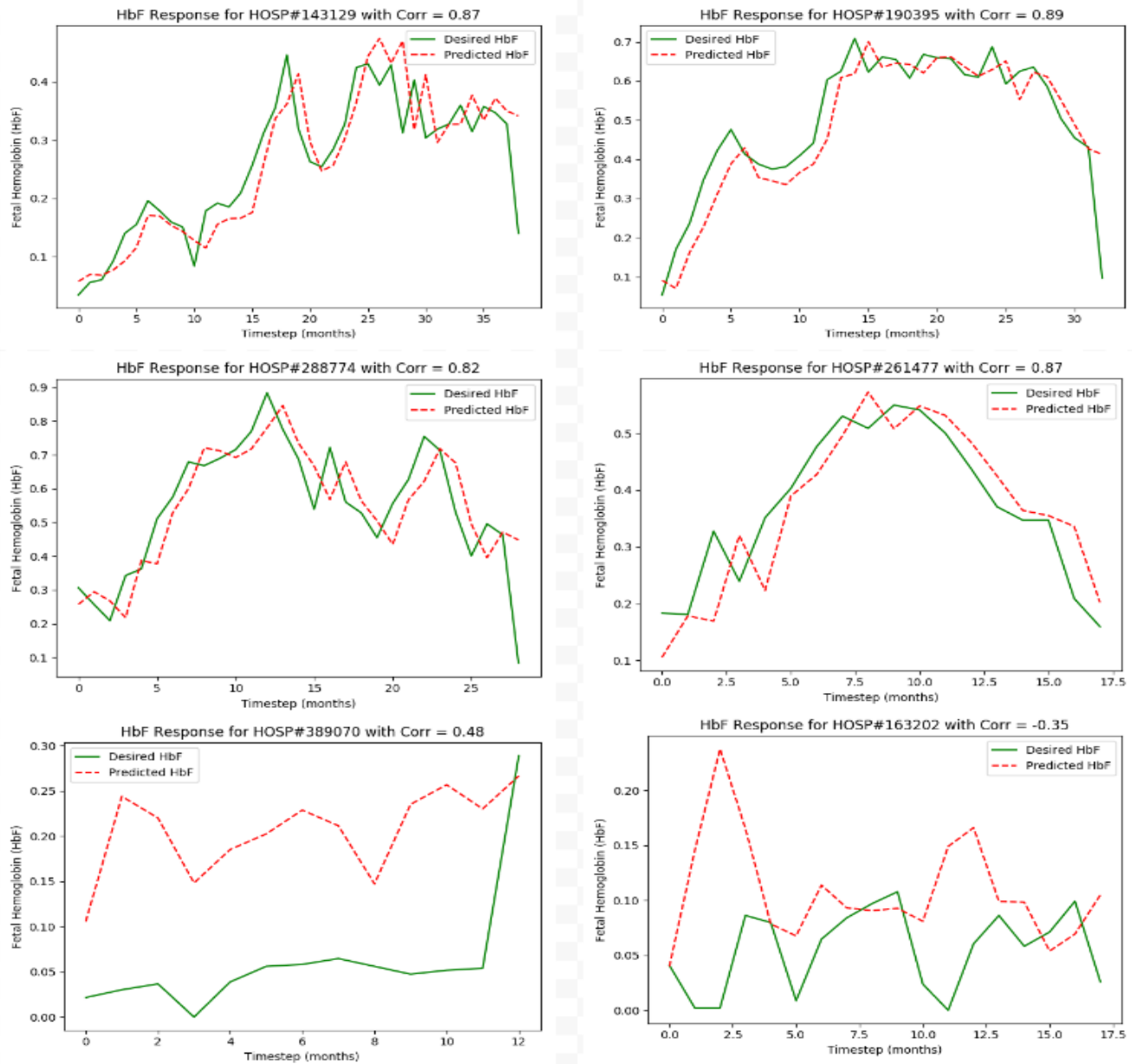

*Figure 9: Some Sample patients time-series results predicted with with the LSTM neural network.(HbF on the Y-axis and Timestep in months on the x-axis)*

# 5. Conclusion

Neural networks are capable of recognizing patterns and complex relationships between parameters which are not easily discernible by man unaided by a machine. The parameter values used for this experiment were obtained before and after the patients underwent Hydroxyurea therapy.

We have demonstrated that we can (with 83% accuracy) identify patients that will respond to Hydroxyurea (HU) therapy. We have been successful at predicting (with 92.6% accuracy) a patient's final fetal hemoglobin value after undergoing HU therapy such that the percentage error between the actual value and the predicted value is less than or equal to 30%. To improve the confidence in the model's prediction, we have gone a step further with an 81.96% accuracy within a 15% error margin of expected and actual HbF values.

Regardless of the way the output classes are categorized, training ANNs to accurately distinguish between different classes of patients is of value to the medical community. This experiment shows that ANNs are capable of exploiting correlations between medical data that physicians are not able to identify unaided. This approach and others like it will positively influence medical decision making, administration of intervention procedures and improve the practice of precision medicine.

We have also demonstrated the potential for using a single, well trained general model to predict the response trajectory of a Sickle Cell Anemia (SCA) patient to Hydroxyurea (HU) with 60% accuracy.